\documentclass[a4paper,twocolumn,showpacs]{revtex4}

\usepackage{graphicx}
\usepackage{dcolumn}
\usepackage{amsmath}
\usepackage{amssymb}
\usepackage{epsfig}

\usepackage{bm}

\begin{document}
\preprint{APS/123-QED}

\title{Fluid pumped by magnetic stress}
\author{Robert Krau\ss, Mario Liu$^1$, Bert Reimann, Reinhard Richter, and
Ingo Rehberg}
\affiliation{Experimentalphysik V, Universit\"at  Bayreuth,  D--95440 Bayreuth,
Germany}
\affiliation{$^1$ Institut f\"ur Theoretische Physik, Universit\"at  T\"ubingen,
D--72076 T\"ubingen, Germany}
\date{\today}

\begin{abstract}
A magnetic field  rotating on the free surface of a ferrofluid
layer is shown to induce considerable fluid motion towards the direction the
field is rolling. The measured flow velocity i) increases with the square of the
magnetic field amplitude, ii) is proportional to the thickness of the fluid layer,
and iii) has a  maximum at a driving frequency of about 3 kHz. The pumping
speed can be estimated with a two-dimensional flow model.  \end{abstract}

\maketitle

Polarizable fluids can show a macroscopic reaction to external electric or
magnetic fields.  While for most fluids the influence of a magnetic field is fairly
weak, colloidal suspensions of magnetic particles -- so called Ferrofluids -- do
show a strong response particularly to static magnetic fields \cite{Rosensweig}.  If
these fields are time dependent, a rich variety of phenomena occurs. The internal
rotation of the magnetization in an externally rotating magnetic field gives rise to
nontrivial effects, as summarized in a recent special  issue \cite {Shliomiseditor}.
A particularly interesting example is the driving of a macroscopic flow by means
of an external rotating magnetic field \cite {Pshenichnikov00}, because it should
allow for a very fine tuning both of the speed and of the direction of the flow even
in microscopic channels \cite {Liu98}.

In this paper we present a novel magnetic fluid \cite {newfluid} in an open flow
geometry especially designed for a
quantitative comparison between the measured flow velocity and its theoretical
estimation.
It is scetched in Fig.\,1.

\begin{figure}[h]
\begin{minipage}{86mm}
\epsfxsize=75mm
\epsfbox{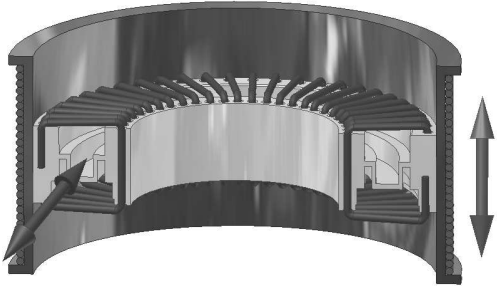}
\end{minipage}
\caption{Experimental setup. The arrows indicate the direction
of the oscillating magnetic fields provided by the coils.}
\label{setup}
\end{figure}
A circular Macrolon$^{\circledR}$ duct with a mean diameter $d$ of 100 mm
and a square cross-section of 5 mm $\times$ 5 mm (and 2.5 mm $\times$ 2.5 mm in
a second set of measurements) is filled brimful with a magnetic fluid
\cite{newfluid} with a viscosity of $\eta=5.4 \cdot 10^{-3}$ Pa\,s. The
orientation of the two coils producing the rotating magnetic field is also indicated
in Fig.\,1: One coil is wrapped around this circular channel and provides a
magnetic field in azimuthal direction, and the outer coil produces the vertical
component of the field. Both coils are driven with an ac-current with a phase
difference of 90$^o$, thus producing a  rotating field on the free surface of the
fluid within the duct.

The characterization of the magnetic susceptibility of the fluid in the duct is of
primary importance for the pumping of the fluid, as discussed in this paper. It has
been measured as a function of frequency of the external oscillating azimuthal
magnetic field by means of a pick-up coil placed into the liquid. Precisely speaking,
the magnetization was determined from the difference of the signal detected by the
pick-up coil in the empty and the filled channel, under the influence of an
oscillatory azimuthal field. The results are presented in Fig.\,2. The measured
susceptibilities of this novel cobalt based fluids are fairly large compared to the
more common magnetite based fluids. In particular, the large imaginary part of the
susceptibility is essential for the pumping action described in this paper.
\begin{figure}[h]
\begin{minipage}{86mm}
\epsfxsize=75mm
\epsfbox{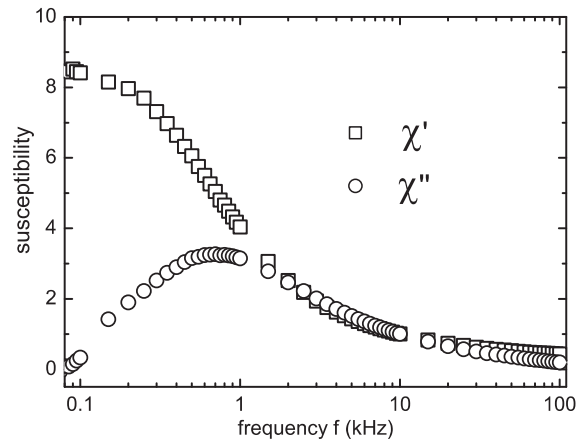}
\end{minipage}

\caption{The magnetic susceptibility as a function of the frequency of the external
alternating magnetic field. Here the real and imaginary parts are denoted by squares and 
circles, respectively.}
\label{susceptibility_vs_f}
\end{figure}

The pump does work: A rotating field produced by the coils leads to a motion of
the fluid in azimuthal direction of the channel \cite{movie}. Its velocity is on the order of mm/s
and can thus be determined by visual inspection of tracer particles swimming on
its surface. The next observation is also a qualitative one: When changing the
phase difference between the ac-fields from +90$^o$ to $-$90$^o$, the flow changes
its sign. The flow direction is such that the vorticity of the flow field
is locally parallel to the rotation vector of the magnetic field, i.\,e.\ the fluid flows
towards the direction the field is rolling.

For quantitative measurements it is necessary to use particles which are small
compared to the channel width (dandruff, diameter about  1 mm).  The velocity is
determined by taking the time a particle  needs to travel a few centimeters. The
size dependence of these measurements is taken into account by using the
numerically calcultated -- roughly parabolic -- velocity profiles , and by
assuming that a floating particle represents the mean speed with respect to
its diameter. A result obtained for a fixed frequency of 1 kHz is presented in Fig.\,3, 
where the maximal flow velocity  within the channel is
shown as a function of the amplitude $G_0$ of the driving external magnetic field.
The velocity increases proportional to $G_0^2$, as demonstrated by the solid
line, a parabola. Here we follow the nomenclature of Ref.\,\cite{Lebedev03}, 
where the external magnetic far field is marked by {$\boldsymbol G$} and the local
one by {$\boldsymbol H$}.
\begin{figure}[h]
\begin{minipage}{86mm}
\epsfxsize=75mm
\epsfbox{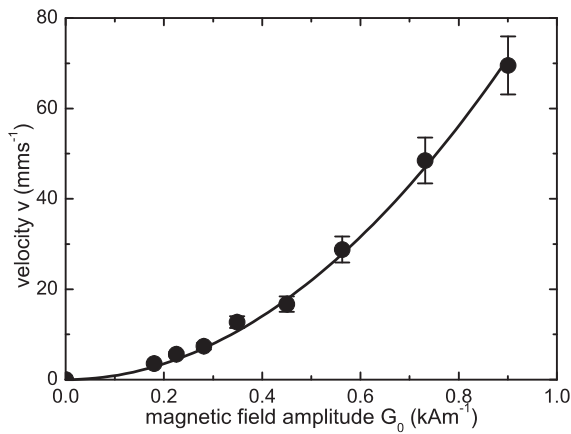}
\end{minipage}
\caption{Maximal velocity measured as a function of the field amplitude at a
fixed frequency of 1\,kHz.}
\label{velocity_vs_H}
\end{figure}

Having demonstrated this quadratic dependence of the velocity on the
magnetic field, a first approach to collapse data obtained at different fields is
to introduce a reduced velocity by dividing with the square of the external field.
Another important influence determining the fluid velocity is the height of the
channel: It turns out that the velocity is larger in bigger channels. This leads us to
reduce the velocity also by dividing by the height $L$ of the duct. In order to
get a dimensionless number one also has to scale with the viscosity of the fluid.
Thus we define
\begin{equation}
u = v_{\rm max}\frac{\eta}{L  \mu_0  G_0^2}
\label{u}
\end{equation}
as a reduced flow velocity. Its measured values are presented as a function of the
driving frequency of the rotating magnetic field in Fig.\,4, for velocities obtained in
a 5 mm $\times$ 5 mm  and a 2.5 mm $\times$ 2.5 mm duct. Both
measurements show a maximum of this velocity in the range of 2 -- 4 kHz.

\begin{figure}[h]
\begin{minipage}{86mm}
\epsfxsize=75mm
\epsfbox{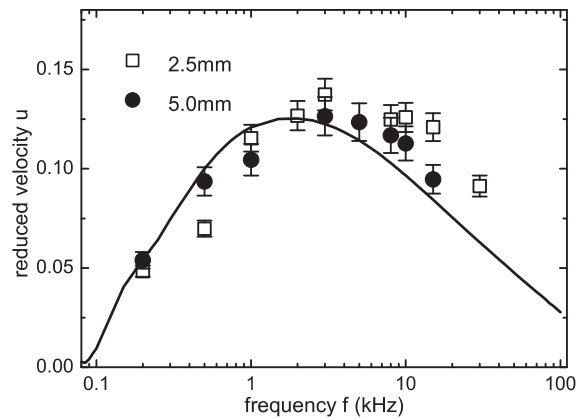}
\end{minipage}
\caption{The reduced velocity $u$ as a
function of the driving frequency. Solid circles (open squares) are obtained in the
5 mm $\times$ 5 mm (2.5 mm $\times$ 2.5 mm) duct, respectively. 
The solid line represents the value expected on
the basis of the measurement of the ac-susceptibility.}
\label{v_vs_f}
\end{figure}

The pumping action can be understood as a manifestation of the magnetic stress
acting on the magnetized fluid, as summarized in Ref.\,\cite{Shliomis03}. This
assumption explains all qualitative features of the
observation: as long as the magnetization is proportional to the magnetic field
(which has been measured to be the case for our fluid, within a precision of
5 \% for  fields up to about 1500 ${\rm Am^{-1}}$), the stress must be proportional to $G^2$
as demonstrated in Fig.\,3. If the
frequency approaches zero, the magnetization and the field are parallel to each
other, the tangential stress is zero and thus the motion of the fluid stops. For finite
frequencies the velocity is proportional to the $\chi\prime\prime$ component of the
susceptibility, which must increase linearly (to lowest order) with the frequency.
For higher frequencies, the imaginary part of the susceptibility $\chi\prime\prime$ has a
maximum at about 1 kHz, which explains that the maximal pumping velocity is
observed around that frequency.

For the quantitative calculation we solve the Laplace equation for the velocity field 
including  the no-slip condition for the fluid at the bottom and the side walls of the duct.
At the upper surface of the fluid the magnetic stress provides the
boundary condition
 \renewcommand{\theequation}{\arabic{equation}}
\begin{equation}
\eta \frac{\partial v}{\partial z}=\frac{\mu _0}{2} ( M_{\rm z} H_{\rm x} -
M_{\rm x} H_{\rm z} ),
\label{magnetic_stress}
\end{equation}
because Eq.\,(65) of Ref. \cite{Shliomis03} is applicable to our geometry.

Both field components $H_{\rm x}$ and $H_{\rm z}$ and the corresponding magnetization
$M_{\rm x}$ and $M_{\rm z}$ are not constant in our case, but depend both on time and space.
By numerical computations of the internal field $\boldsymbol H$ and the resulting magnetization 
$\boldsymbol M=\chi \boldsymbol H$, which will be published elsewhere, we get for the maximum
fluid velocity in the middle of the upper surface 
\begin{equation}
v_{\rm max}= \chi\prime\prime \frac{\mu_0  G_0^2}{2}
\frac{L} {\eta}  \alpha \left(\frac{1+N_{\rm eff} \chi\prime}{(1+N_{\rm eff}\chi\prime )^2+
(N_{\rm eff}\chi\prime\prime )^2}\right).
\label{vmax}
\end{equation}
Here $N_{\rm eff}$ denotes an effective demagnetization factor and $\alpha$ a reduction factor 
due to the given geometry depending on the aspect ratio   $a=L_{\rm y}/L_{\rm z}$ of the
rectangular channel. In our case $a=1$ and we obtain $\alpha \approx 0.369$.

In our square cross section, the magnetization -- and  the ensuing stress at the surface of
the fluid -- is not homogenous. We thus extract an
effective demagnetization factor $N_{\rm eff}$ by fitting
Eq.\,(\ref{vmax}) to the numerically obtained velocity for different values of $\chi$. We
obtain $N_{\rm eff} \approx 0.656$, which seems realistic when comparing to $N=0.5$
for the case of a circular cylinder.

From Eqs.\,(\ref{u},\ref{vmax}) we finally get the theoretical estimation for the reduced velocity
\begin{equation}
u=\frac{\alpha}{2} \frac{\chi\prime\prime
(1+N_{\rm eff}\chi\prime) }{(1+N_{\rm eff}\chi\prime)^2+(N_{\rm eff}\chi\prime\prime)^2}
\label{u_2}
\end{equation}
The limiting case of an infinitly wide channel
($N=1, \alpha=1$) is included in this formula.

The reduced velocity obtained from Eq.\,(\ref{u_2}) is presented in Fig.\,4 as a solid line,
with the values of $\chi\prime$ and $\chi\prime\prime$ taken from the
measurements presented in Fig.\,2. The agreement between the measured
velocities and the solid line is on a 20\% level. This can partly be attributed to the
limited accuracy of the measurement procedure, which is indicated by the error
bars in Fig.\,4. The systematic deviations between the data and the theoretical
curve are believed to reflect the precision of the simplifying assumptions going
into the consideration presented above. For example, the influence of the shape of
the meniscus of the fluid adds a three-dimensional complication to the problem,
whose influence on the maximal pumping speed is hard to estimate. Moreover, it
should be noted that magnetic fluids are not perfectly stable both in their
magnetic susceptibility and their viscosity, which might add to the small mismatch
between the expectation based on the measurement of the susceptibility and the
measured velocity. Finally,  taking into account the small amplitude of the magnetic field,
any magnetoviscous effects have been neglected for the calculations.

 In summary, the pump presented here works well and has an interesting potential
 especially in small geometries where a mechanical driving of the flow is not
 possible. More importantly, it does seem safe to conclude that the ansatz of a
 magnetic  stress driven motion captures the essence of this pump on a
 quantitative level.

It is a pleasure to thank A.~Engel, M.~Krekhova and M.\,I.~Shliomis for clarifying
discussions. We thank N.~Matoussevitch for providing the magnetic liquid. The
experiments were supported by Deutsche Forschungsgemeinschaft, Re 588/12.

\end{document}